\documentclass[preprint,aps,showpacs,nofootinbib]{revtex4}

\usepackage{amssymb}
\usepackage{bm}

\def\bea{\begin{eqnarray}}
\def\eea{\end{eqnarray}}
\def\bec{\begin{center}}
\def\ec{\end{center}}

\def\beq{\begin{equation}}
\def\eeq{\end{equation}}

\begin{document}

\begin{center}
{\Large \bf  Sparticle Spectrum of Large
Volume Compactification}
\end{center}

 \vspace*{5mm} \noindent
\centerline{\bf Kiwoon Choi${}^{a}$,    Hans Peter Nilles${}^{b}$,
Chang Sub Shin${}^{a}$, Michele Trapletti${}^{b}$} \vskip 1cm
\centerline{\em ${{}^{a}}$ Department of Physics, Korea Advanced
Institute of Science and Technology} \centerline{\em Daejeon
305-701, Korea} \vskip 5mm \centerline{\em ${{}^{b}}$ Bethe Center for Theoretical Physics and Physikalisches
Institut, Universit\"at Bonn} \centerline{\em Nussallee 12, D-53115
Bonn, Germany}

\vskip .9cm

\centerline{\bf Abstract} \vskip .3cm

We examine the large volume compactification of Type IIB string
theory or its $F$ theory limit and the associated supersymmetry
breakdown and soft terms. It is crucial to incorporate the
loop-induced moduli mixing, originating from radiative corrections
to the K\"ahler potential. We show that in the presence of moduli
mixing, soft scalar masses generically receive a $D$-term
contribution of the order of the gravitino mass $m_{3/2}$ when the
visible sector cycle is stabilized by the $D$-term potential of an
anomalous $U(1)$ gauge symmetry, while the moduli-mediated gaugino
masses and $A$-parameters tend to be of the order of
$m_{3/2}/8\pi^2$. It is noticed also that a too large moduli mixing
can destabilize   the large volume solution by making it a saddle
point.


\vskip .3cm


\newpage

\section{introduction}

Supersymmetry (SUSY) breakdown in string theory often requires the
presence of fluxes and nonperturbative effects (like gaugino
condensation) \cite{DIN,GKP}. The complete picture, however,
requires the fixing of many moduli fields. This might be relatively
easy to fulfill in the framework of type IIB theory. Three-form
fluxes, gaugino condensation (or $D$-brane instantons) and a
specific uplifting mechanism \cite{KKLT} lead to a picture where the
gravitino mass and the soft supersymmetry breaking terms can be
determined explicitly \cite{mirage1,mirage1a}. The smallness of the
gravitino mass (compared to the string and Planck scales) in the low
energy effective supergravity theory originates from a small
constant term $W_0$ in the superpotential\footnote{In this picture,
the uplifting potential is exponentially small as it arises from a
SUSY breakdown at the tip of warped throat, and then small $W_0$ is
required to tune the cosmological constant to a nearly vanishing
value.}. The mediation of SUSY breakdown is a variant of gravity
mediation. In the simplest case the contribution to the soft terms
is suppressed by the factor $1/\ln(M_{\rm Planck}/m_{3/2})$ compared
to the gravitino mass $m_{3/2}$ and radiative corrections such as
anomaly mediation become competitive, leading to a scheme called
mirage mediation \cite{mirage1a,mirage2}. Then the scale of soft
terms is set essentially by $m_{3/2}$, suppressed by a factor
$1/\ln(M_{\rm Planck}/m_{3/2})\sim 1/4\pi^2$, and the gravitino mass
should thus be in the multi-TeV range.

Recently, alternative metastable local minima of the scalar
potential in type IIB theory have been analyzed. They are
characterized by  large compactification volume,
leading to the so-called large volume scenario (LVS) \cite{LVS}. An
attractive feature of this scenario is the fact that a small value of
$m_{3/2}$ does not require a small constant $W_0$ in the
superpotential, but can rather be the consequence of large volume
suppression. With $W_0\sim{\cal O}(1)$, we have the approximate
relation $M_{\rm Planck}/M_{\rm string}\sim M_{\rm string}/m_{3/2}$.
Values of $m_{3/2}$ in the TeV range would then require the string
scale at an intermediate scale around $10^{11}$ GeV. Therefore it
might be difficult to incorporate the idea of grand unification (at
a scale around $10^{16}$ GeV) in the large volume scenario.

Of course, a full understanding of the situation needs a reliable
computation of soft SUSY breaking terms and their relation to the
gravitino mass, which is not available yet. In a recent paper
\cite{Blumenhagen et al.}, it has been suggested that the soft terms
might be tiny compared to $m_{3/2}$, for instance $m_{\rm soft}\sim
m_{3/2}^{3/2}/M_{\rm Planck}^{1/2}$ or even as small as $m_{\rm
soft}\sim m_{3/2}^2/M_{\rm Planck}$. This would then allow the
gravitino mass as large as $10^{11}$ GeV and the string scale to be
of order of $10^{15}$ GeV, which would allow the incorporation of
gauge coupling unification at $M_{GUT}\sim 10^{16}$ GeV
\cite{Blumenhagen et al.} in the LVS scheme. In a more recent paper
\cite{Conlon-Pedro}, however, it has been argued that a mechanism of
``moduli mixing" could destabilize this hierarchy and bring the soft
terms closer to the gravitino mass. This is reminiscent of similar
discussions in heterotic string theory (as well as Horava-Witten
theory), where higher order corrections to gauge kinetic functions
and K\"ahler potential were shown to destabilize such hierarchies
between $m_{\rm soft}$ and $m_{3/2}$ \cite{heterotic,horava-witten}.
As was pointed out in \cite{Blumenhagen:2007sm}, it is difficult to
generate a nonperturbative superpotential for the visible sector
4-cycle supporting chiral matter fields, which points toward a
$D$-term stabilization of the visible sector 4-cycle in LVS
\cite{Blumenhagen et al.}. Then there might be additional
contribution to soft terms arising from the $D$-term scalar
potential.

In this paper we would like to analyze the large volume
compactification and the associated soft terms in a class of
LVS-theories taking into account the potential ``instabilities" of
this scenario. Radiative corrections to the K\"ahler potential and
the resulting moduli mixings are shown to become important for the
values of soft terms, completely dominating the contributions of
order $m_{\rm soft}\sim m_{3/2}^{3/2}/M^{1/2}_{\rm Planck}$
discussed in \cite{Blumenhagen et al.}.
 We also stress the importance of the $D$-terms along the
visible sector 4-cycle in the LVS-models. They tend to dominate the
soft scalar mass terms and give a contribution of the order of the
gravitino mass $m_{3/2}$. Gaugino masses and A-parameters do not
receive $D$-term contributions, and generically tend to be
loop-suppressed, being  of the order of ${\cal O}(m_{3/2}/8\pi^2)$.
With these contributions from moduli mixings, if the gravitino mass
were of the order of $10^{11}$ GeV as conjectured in
\cite{Blumenhagen et al.} (in order to accommodate the unification
scale $M_{GUT}\sim 10^{16}$ GeV with $W_0\sim {\cal O}(1)$), severe
fine-tuning of the K\"ahler potential  at the multi-loop level would
be required to keep the soft terms in the TeV range. We would
therefore argue that the gravitino mass should not exceed the
(multi)-TeV range.

This paper is organized as follows. In section 2, we revisit the
LVS-scheme while incorporating the moduli redefinition discussed in
\cite{Conlon-Pedro}. We also include a discussion of the stability
of the large volume solution in the presence of moduli mixing.
Section 3 discusses the $D$-term stabilization of the visible sector
cycle with an explicit scheme to stabilize the remained $D$-flat
direction which is parameterized  in this case by $U(1)_A$-charged
(but MSSM singlet) matter fields breaking a global Peccei-Quinn
symmetry spontaneously. This scheme naturally generates an
intermediate axion scale in LVS, and can be implemented in other
scenarii with a high string scale close to the Planck scale.
Section 4 is devoted to the discussion of supersymmtry breakdown and
resulting soft terms, and conclusions and outlook will be given in
section 5.

\section{Large volume compactification with moduli mixing}

In this section, we revisit the large volume scenario (LVS) while
incorporating the loop-induced moduli mixing discussed in
\cite{Conlon-Pedro}. To achieve a large compactification volume, one
needs at least two K\"ahler moduli, $T_1$ and $T_2$, where $T_1$
describes a 4-cycle with large volume and $T_2$ stands for a small
4-cycle supporting non-perturbative effects such as $D3$ instantons.
An exponentially large vacuum value of $t_1=T_1+T_1^*$ is obtained
by the competition between the $\alpha^\prime$-correction of ${\cal
O}(1/t_1^{3/2})$ and the $D3$ instanton effect of ${\cal
O}(e^{-aT_2})$, yielding $t_1^{3/2}\sim |e^{a T_2}|$. In 4D
effective SUGRA of LVS, the 4D Planck scale $M_{\rm Planck}\equiv
1/\sqrt{8\pi G_N}$ and the cutoff scale $\Lambda$ of local dynamics
on small 4-cycle differ by certain powers of the compactification
volume. For instance, in type IIB theory we have \bea
\label{string-scale} \frac{M^2_{\rm Planck}} {M^2_{\rm
string}}\,\sim\, {\cal V}_{CY}\,\sim\,t_1^{3/2},\eea where ${\cal
V}_{CY}$ denotes the Calabi-Yau volume in the string length unit
with $M_{\rm string}=1$, and therefore $\Lambda/M_{\rm Planck}\sim
1/t_1^{3/4}$ for the local cutoff scale $\Lambda\sim M_{\rm
string}$.  Generically there can  be radiative corrections
localized on a small 4-cycle, which are controlled by a coupling
inversely proportional to the 4-cycle volume $t_2=T_2+T_2^*$, and
also depend logarithmically on the local cutoff scale $\Lambda$.
Since $M_{\rm Planck}$ is the natural mass scale of 4D effective
SUGRA, including those quantum corrections in the K\"ahler potential
and/or the gauge kinetic functions would require a redefinition of
$t_2$, involving $\ln (M_{\rm Planck}/\Lambda)\propto \ln t_1$
\cite{Conlon-Pedro} as \bea \label{redef_IIB} t_2\,\rightarrow\,
\tilde{t}_2 =t_2 -\alpha_2\ln t_1,\eea where $\alpha_2$ is a
parameter representing the size of the quantum corrections that lead
to the above moduli-mixing. (Note that this is not a redefinition of
the chiral superfield, but a redefinition of the scalar component of
the chiral superfield.) In fact, a similar phenomenon has been noticed
in heterotic string/$M$ theory context, i.e. a redefinition of the
heterotic string dilaton \cite{moduli-mixing}\bea
\label{redef_heterotic}s\rightarrow s-\alpha \ln t,\eea which is
required to accommodate the loop threshold correction to 4D gauge
coupling \cite{gauge-threshold} where $s=S+S^*$ is the heterotic
string dilaton and $t=T+T^*$ is a K\"ahler modulus in underlying
heterotic string compactification. In the heterotic $M$-theory
limit, $s$ corresponds to the small volume of 6D internal space,
while $t$ describes the large length of the 11-th dimension. Then
the heterotic redefinition (\ref{redef_heterotic}) in the limit
$t\gg s$ can have a similar geometric interpretation as the type IIB
field redefinition (\ref{redef_IIB}) in LVS.

To proceed,  following \cite{LVS, Blumenhagen et al.},  we  assume that
all complex structure moduli and the string dilaton  are stabilized
by fluxes at a supersymmetric solution, and those flux-stabilized
moduli can be integrated out without affecting  the subsequent
stabilization of K\"ahler moduli.\footnote{Once $T_i$ are stabilized
at the SUSY-breaking vacuum, nonzero $F$-components of the dilaton and
complex structure moduli do appear as well. However their effects
are subleading compared to those of $F^{T_i}$ in the large volume
limit $t_1\gg 1$.} In the following, unless specified, we set the 4D
Planck scale (in the Einstein frame) $M_{\rm Planck}=1$.

Since we are interested in the large volume limit $t_1\gg 1$, it is
convenient to expand the K\"ahler potential of the model in
(appropriate) powers of $1/t_1$.
Then, after the dilaton and complex structure moduli are integrated
out, the K\"ahler potential and superpotential of $T_i$ ($i=1,2$)
are given by \cite{Conlon-Pedro}\bea \label{4dsugra} K&=&-3\ln
t_1+\frac{2(\tilde{t}_2^{3/2}-\xi_{\alpha^\prime})}{t_1^{3/2}}+{\cal
O}\left({1}/{t_1^3}\right), \nonumber
\\
W&=&W_0+Ae^{-aT_2},\eea where \bea \label{t2mixing}
\tilde{t}_2=t_2-\alpha_2\ln t_1 \quad (t_i=T_i+T_i^*).
 \eea
  The parameter
$\xi_{\alpha^\prime}$ in the K\"ahler potential  represents the
$\alpha^\prime$-correction,
 and $\alpha_2$ parameterizes  the radiative corrections that lead
to the loop-induced redefinition  of $t_2$.
 Here we assume
that $Ae^{-aT_2}$ is induced by $D3$ instantons, so \bea
\frac{at_2}{2}=\mbox{Euclidean action of $D3$ instanton}.\eea The
constant $W_0$ in the superpotential might arise from 3-form fluxes,
and is assumed to be of order unity in LVS. As we will see, for the
model of (\ref{4dsugra}), we have \bea \label{gravitino-mass}
\frac{m_{3/2}}{M_{\rm Planck}}\,=\,e^{K/2}|W|\,\sim\,
\frac{|W_0|}{t_1^{3/2}} \,\sim\, |Ae^{-aT_2}|\eea regardless of the
value of $W_0$. As a result,  the value of $D3$ instanton action is
given by \bea \label{d3action} \frac{at_2}{2}\,\sim\, \ln (M_{\rm
Planck}/m_{3/2})\,\sim\, 4\pi^2 \eea regardless of whether $W_0\sim
{\cal O}(1)$ or hierarchically small. On the other hand, with Eq.
(\ref{t2mixing}), $\alpha_2/t_2$ corresponds to a loop suppression
factor on small 4-cycle,  and thus is expected to be ${\cal
O}(1/8\pi^2)$, which implies  \bea a\alpha_2 = {\cal O}(1).\eea
 With appropriate
$R$-transformation and axionic shift ${\rm Im}(T_2)\rightarrow {\rm
Im}(T_2)+$ constant, we can always make $W_0$ and $A$ to be real
positive parameters. In the following, we will take such a field
basis in which $W_0$ and $A$ are real and positive.

Let us now examine the stabilization of K\"ahler moduli in the 4D
effective  SUGRA model of (\ref{4dsugra}).
The scalar potential of the model is given by \bea V_{\rm
SUGRA} = e^K\left\{K^{I\bar J} D_I W (D_{J}W)^* - 3|W|^2\right\}.
 \eea
Assuming  $t_1\gg 1$ and also $a\tilde{t}_2={\cal O}(8\pi^2)$ as
indicated  by (\ref{d3action}), we can solve the equations of motion
\bea
 \partial_{t_1}V_{\rm SUGRA}=\partial_{t_2}V_{\rm SUGRA}=0\eea
 to find  \bea \label{lvs-sol}&&
 \frac{t_1^{3/2}}{W_0} =
\frac{e^{at_2/2}}{aA}\xi_{\alpha'}^{1/3}
\left(\frac{3}{2}-\frac{21+8 a\alpha_2 }{12 a\tilde t_2}+{\cal
O}\left(\frac{1}{(a\tilde t_2)^{2}}\right)\right),
\nonumber\\
&& \tilde t_2^{3/2} = \xi_{\alpha'} \left(1 + \frac{3-13
a\alpha_2}{3a\tilde t_2} + {\cal O}\left(\frac{1}{(a\tilde
t_2)^{2}}\right)\right),
 \eea
 where the solution is expanded in powers of $1/a\tilde{t}_2$.
 The above solution shows that
  $t_1$ can indeed have an exponentially large
 value
for the parameter values given that $at_2\gg 1$. Using the above
solution, we can find the following moduli mass spectrum: \bea
\label{moduli-mass} && m_{t_1}^2 \,\approx\,
(3-a\alpha_2)\frac{\xi_{\alpha'}}{a\tilde t_2}
\frac{m_{3/2}^2}{t_1^{3/2}},\quad m_{a_1}^2 \,=\,0,\nonumber
\\
&&  m_{t_2}^2\,\approx\, m_{a_2}^2\,\approx\, (a\tilde t_2)^2
m_{3/2}^2,\eea implying that we need \bea
\label{vacuum-sta}a\alpha_2 < 3\eea in order for the solution
(\ref{lvs-sol}) to be a (meta) stable local minimum.

On the other hand, the form of the K\"ahler potential and
superpotential in (\ref{4dsugra}) is valid only when both $t_2$ and
$\tilde{t}_2$ are positive in the convention that $a$ is positive,
which corresponds to the condition that both the $D3$ instanton
action and the moduli K\"ahler metric are positive. Combined with
the vacuum solution (\ref{lvs-sol}), this leads to \bea
\label{consistency} \frac{3\tilde{t}_2}{t_2}=
3-a\alpha_2+\frac{2\alpha_2}{t_2}\Big(\ln (1/w_0) +{\cal O}(1)\Big)
\,>\, 0.\eea For the case with $W_0\sim {\cal O}(1)$, since
$\alpha_2/t_2 ={\cal O}(1/8\pi^2)$,  the vacuum stability condition
(\ref{vacuum-sta}) is satisfied for most part of the parameter
region satisfying the above condition. Although $W_0$ is assumed  to
be ${\cal O}(1)$ in usual LVS, it is often required to be
hierarchically small in order to accommodate both the gauge
unification at $10^{16}$ GeV and the soft SUSY breaking masses $\sim
1$ TeV. For instance, in the presence of moduli mixing, our
discussion in the next section implies that sfermions in the visible
sector can get a soft mass of ${\cal O}(m_{3/2})$ from the $D$-term
contribution, while the gaugino masses are of ${\cal
O}(m_{3/2}/8\pi^2)$. In this case, demanding gauge unification
at $10^{16}$ GeV and the weak scale size of gaugino masses, one
finds $M_{\rm string}\sim 10^{15}$ GeV and $m_{3/2}\sim 10$ TeV,
which correspond to $t_1\sim 10^3$ and $W_0\sim 10^{-10}$ (see
(\ref{string-scale}) and (\ref{gravitino-mass})).
 In such case of small $W_0$, there can be a sizable
part of parameter space which satisfies (\ref{consistency}), but is
excluded by the stability condition (\ref{vacuum-sta}).

For $a\alpha_2<3$, the large volume solution (\ref{lvs-sol})
corresponds to an AdS vacuum (at least at tree level) as it gives
the vacuum energy density\bea \label{ads} V_{\rm vacuum}\approx -
(3-a\alpha_2)\frac{\xi_{\alpha'}}{a\tilde
t_2}\frac{m_{3/2}^2}{t_1^{3/2}}. \eea Then, to achieve a
phenomenologically viable de-Sitter vacuum, one might need an
uplifting potential induced by additional sources of SUSY
breaking\footnote{Alternatively, one can assume that (quadratically
divergent) radiative corrections to the vacuum energy density do the
job of uplifting \cite{Choi-Kim-Nilles}.}. Since $m_{t_2}={\cal
O}(8\pi^2 m_{3/2})$ as indicated by (\ref{moduli-mass}), such an
uplifting potential would not significantly affect the vacuum
solution along the $t_2$-direction, however it can cause a large
shift of $t_1$, and even might destabilize the solution.

It is unclear yet if a successful uplifting sector can be introduced
within the framework of large volume compactification of  type IIB
theory or its $F$-theory limit\footnote{In KKLT compactification
with warped throat, a successful uplifting can be achieved by
anti-$D3$ branes (or any SUSY breaking branes) stabilized at the tip
of throat.}. Here we take a phenomenological approach, simply
introducing an uplifting potential inversely proportional to certain
powers of the CY volume ${\cal V}_{CY}$: \bea \Delta V_{\rm
lift}=\frac{D_0}{{\cal
V}^{n_0}_{CY}}=\frac{D_0}{t_1^{3n_0/2}}\left(1+{\cal
O}\left(\frac{1}{t_1^{3/2}}\right)\right),\eea where $n_0$ is a
positive rational number and $D_0$ is a positive constant which
should be tuned to make the final vacuum energy density nearly zero.
We then find that the stationary solution of the total potential
$V_{\rm TOT}=V_{\rm SUGRA}+\Delta V_{\rm lift}$ is given by
 \bea\label{sol-uplifting} &&
\frac{t_1^{3/2}}{W_0} = \frac{e^{at_2/2}}{aA}\xi_{\alpha'}^{1/3}
\left(\frac{3}{2}-\frac{(21-9n_0) +(8-2n_0) a\alpha_2 }{4(3-n_0)
a\tilde t_2} +{\cal O}((a\tilde t_2)^{-2})\right)\nonumber\\
&&
 \tilde
t_2^{3/2} = \xi_{\alpha'} \left(1 + \frac{3-(13-4n_0)
a\alpha_2}{(3-n_0)a\tilde t_2} + {\cal O}((a\tilde
t_2)^{-2})\right), \eea showing that the qualitative feature of
moduli vacuum values is untouched by the uplifting potential.
Considering the moduli masses, we find $m_{t_1}$ is modified \bea
m_{t_1}^2 = (3-n_0)(3-a\alpha_2)\frac{\xi_{\alpha'}}{t_1^{3/2}}
\frac{m_{3/2}^2}{ a \tilde t_2} \eea while the other moduli masses
are the same as those in (\ref{moduli-mass}). Since the uplifting is
required only for $a\alpha_2<3$ (see the vacuum energy density
(\ref{ads})), this form of the large volume modulus mass implies
that the stationary solution (\ref{sol-uplifting}) becomes a saddle
point when $n_0>3$, for which  the uplifting potential is too steep
to give a local de-Sitter minimum.

\section{$D$-term stabilization of the MSSM cycle modulus}

It has been pointed out in \cite{Blumenhagen:2007sm} that a 4-cycle
supporting chiral matter fields can not have a $D3$ instanton
superpotential. This implies that the $D3$-instanton 4-cycle
described by the K\"ahler modulus $T_2$ can not be identified as the
visible sector 4-cycle supporting the MSSM gauge and matter fields.
The large volume 4-cycle also can not be identified as the visible
sector cycle as it would give a too small SM gauge coupling
$g_{SM}^2\sim 1/t_1$. With this observation, a third 4-cycle
 has been introduced  in \cite{Blumenhagen et al.} to
accommodate  the MSSM sector
  under the assumption  that the corresponding K\"ahler modulus
$T_3$ is stabilized by a $D$-term potential.

As there is no instanton superpotential of the form $e^{-bT_3}$, the
$D$-term stabilization of $T_3$ is indeed a natural direction to be
explored. For this, we need a $D$-term which depends on
$t_3=T_3+T_3^*$ even in the limit that other (gauge charged) matter
fields are all vanishing. This would be achieved by having an
anomalous $U(1)$ symmetry under which $T_3$ transforms nonlinearly
\cite{u1a}, so that the associated anomaly is canceled by the
Green-Schwarz (GS) mechanism \cite{Green-Schwarz}: \bea U(1)_A:\quad
V_A\rightarrow V_A+\Lambda_A+\Lambda_A^*, \quad T_3\rightarrow
T_3+\delta_{GS}\Lambda_A, \quad \Phi_i\rightarrow
e^{-2q_i\Lambda_A}\Phi_i, \eea where $V_A$ is the vector superfield
containing the $U(1)_A$ gauge field, $\Lambda_A$ is a chiral
superfield parameterizing the $U(1)_A$ transformation on $N=1$
superspace, $\delta_{GS}$ is a constant of ${\cal O}(1/8\pi^2)$
(under a proper normalization of $T_3$), and finally $\Phi_i$ stand
for generic chiral matter superfields with $U(1)_A$ charge $q_i$.
 The gauge boson mass and the $D$-term of  $U(1)_A$
are given by  \bea \label{u1mass} M_A^2 &=& 2g_A^2 \eta^I\eta^{\bar
J} K_{I\bar J} \,\equiv\,
2g_A^2\Big(M_{GS}^2 + M_{PQ}^2\Big), \nonumber \\
D_A &=& -\eta^I K_I \,\equiv\,  \xi_{FI} +\tilde{M}_{PQ}^2, \eea
where $g_A$ is the $U(1)_A$ gauge coupling,
$2\eta^I=-\delta\Phi^I/\delta\Lambda_A=\{-\delta_{GS},2q_i\Phi_i\}$
denote the holomorphic Killing vector field generating an
infinitesimal $U(1)_A$ transformation of  chiral superfields
$\Phi^I=\{T_3,\Phi_i\}$, and \bea \label{u1mass1} {M^2_{GS}}
&\equiv&
\left(\frac{\delta_{GS}}{2}\right)^2\left.\left\langle\frac{\partial^2K}{\partial
T_3\partial T_3^*}\right\rangle\right|_{\Phi_i=0},\nonumber
\\
 {\xi_{FI}}&\equiv&
\frac{\delta_{GS}}{2}\left.\left\langle\frac{\partial
K}{\partial{T_3}}\right\rangle\right|_{\Phi_i=0}. \eea Note that
$M_{GS}^2$ corresponds to the $U(1)_A$ gauge boson mass-square in
the limit $\Phi_i=0$, which originates from the St\"uckelberg
mechanism associated with the $U(1)_A$ transformation of ${\rm
Im}(T_3)$, while $M_{PQ}^2\propto \langle\Phi^*_i\Phi_i\rangle$
stands for the contribution to $M_A^2$ from the nonzero vacuum
values of the $U(1)_A$ charged matter fields $\Phi_i$. (Here we use
the subscript ``PQ"  as $M_{PQ}$ corresponds to the scale of
spontaneous breaking of a $U(1)$ Peccei-Quinn   symmetry \cite{pq}.)
The nonlinear transformation of $T_3$ under $U(1)_A$ gives rise to
the moduli-dependent Fayet-Illiopoulos (FI) term $\xi_{FI}$ which
should be canceled by the $D$-term contribution $\tilde{M}_{PQ}^2$
from the vacuum values of  $\Phi_i$.

The explicit realization of the $D$-term stabilization of $T_3$
depends on the relative size of $\xi_{FI}$ compared to $M_{GS}^2$
\cite{arkani,Choi-Jeong}, which would identify the $D$-flat
direction. If $T_3$ is stabilized at a point with $|\xi_{FI}|\gg
M_{GS}^2$,
 the
$U(1)_A$ gauge boson gets most of its mass from the vacuum values of
$\Phi_i$ since $\tilde{M}_{PQ}^2\approx -\xi_{FI}$ along the
$D$-flat direction and also generically $|M_{PQ}^2|\sim
|\tilde{M}_{PQ}^2|$. In this case, the $D$-term potential fixes
essentially a combination of $U(1)_A$-charged matter fields
$\Phi_i$, while leaving $T_3$ as an unfixed $D$-flat direction.
This would be the case when $T_3$ is stabilized at a geometric
regime where $t_3\sim 1/g_{GUT}^2={\cal O}(1)$ and the K\"ahler
potential
 obeys a simple scaling behavior $\partial K/\partial t_3\sim
t_3\partial^2 K/\partial^2 t_3$, giving
\bea\frac{\xi_{FI}}{M_{GS}^2}\,=\,
\frac{2}{\delta_{GS}}\left.\left\langle\frac{\partial_{T_3}K}{\partial_{T_3}\partial_{T_3^*}K}\right\rangle\right|_{\Phi_i=0}
\,\sim\, \frac{t_3}{\delta_{GS}}\,\sim\, {\cal O}(8\pi^2).\eea


Alternatively,  $T_3$ might be stabilized at or near a (singular)
point with vanishing FI term \cite{Blumenhagen et al.,Conlon-Pedro}.
In such case,
  it is possible to have \bea
  \xi_{FI}\,\sim\, \tilde{M}_{PQ}^2\,\sim\, M_{PQ}^2 \,\ll\,
  M_{GS}^2,\eea
  so that the $U(1)_A$ gauge boson mass-square is dominated by  the St\"uckelberg contribution
  $M_{GS}^2$.
Then  $t_3$ is stabilized as desired by the $D$-term potential of
$U(1)_A$, while the unfixed $D$-flat direction is described mostly
by a combination of $\Phi_i$. This limit  is particularly
interesting since it offers the possibility that $M_{PQ}$ is far
below the string scale. As we will see, in this case, $M_{PQ}$ can
be identified as the scale of spontaneous breakdown of an anomalous
global symmetry which can solve the strong CP problem
\cite{pq,axion1,axion2}. One can then have a QCD axion  with
phenomenologically favored decay constant $M_{PQ}\sim 10^9-10^{12}$
GeV \cite{axion3} even when the string scale is close to $M_{\rm
Planck}$. Another interesting feature of this case is that the
$D$-flat direction can be easily stabilized by the combined effects
of supersymmetry breaking and appropriate tree-level superpotential
of $\Phi_i$. Thus, in the following, we will focus on the case with
$\xi_{FI}\ll M_{GS}^2$.


 To proceed, let $V_a$  denote the MSSM gauge
superfields localized on $D$-branes wrapping the visible sector
4-cycle.
Including $V_A$ and the chiral matter fields $\Phi_i$ together, and
keeping only the leading order terms in the expansion in $1/t_1$,
the K\"ahler potential, superpotential and gauge kinetic functions
of the model are given by
 \bea \label{model}  K&=& K_0(t_1,t_2,t_3)+Z_i\Phi_i^*e^{2q_iV_A}\Phi_i+{\cal O}(|\Phi|^4),\nonumber
\\
W&=&\tilde W_0(T_I)+\frac{1}{3!}\lambda_{ijk}\Phi_i\Phi_j\Phi_k +
{\cal O}(\Phi^4),
\nonumber \\
f_a&=& \gamma_a + k_a T_3, \qquad f_A\,=\, \gamma_A +k_A T_3,\eea
where \bea \label{model-kahler}K_0 &=& -3\ln
t_1+\frac{2(\tilde{t}_2^{3/2}-\xi_{\alpha^\prime})}{t_1^{3/2}}
+\frac{1}{2t_1^p}\Big(\tilde{t}_3^2+{\cal
O}(\tilde{t}_3^3)\Big),\nonumber \\
\tilde W_0&=& W_0+Ae^{-aT_2} \eea for
\bea && t_i=T_i+T_i^* \quad (i=1,2,3), \nonumber \\
&& \tilde{t}_2=t_2-\alpha_2\ln t_1, \quad
\tilde{t}_3=t_3-\alpha_3\ln t_1 -\delta_{GS}V_A,
 \eea
 and $f_a$ and $f_A$ denote the holomorphic gauge kinetic functions
 for the MSSM and $U(1)_A$ gauge
 superfields, respectively.
The constants $\gamma_a$ and $\gamma_A$ in the gauge kinetic
functions might be induced by the vacuum value of the type IIB
string dilaton, and they have a value of order unity (or smaller) in
general. As was noticed in
\cite{Conlon-Pedro,moduli-mixing}, generically it
is expected that the radiative corrections on the visible sector
4-cycle require a redefinition of $t_3$
 depending on  $\ln (M_{\rm Planck}/\Lambda)\propto \ln t_1$, where $\Lambda$ is the local cutoff
scale : \bea t_3 \,\rightarrow\, t_3-\alpha_3\ln t_1.\eea

The form of the K\"ahler potential of $t_3$ is dictated by the
condition that $t_3$ is stabilized near the point
  with vanishing FI-term, which can be defined as
  \bea\left.\frac{\partial K}{\partial
\tilde{t}_3}\right|_{\tilde{t}_3=0}=0\eea
for the $U(1)_A$-invariant combination $\tilde{t}_3$ including the
effect of moduli mixing. For simplicity, here we assume that the
K\"ahler metric of $t_3$ is independent of the second K\"ahler
modulus $t_2$, however all of our subsequent discussions are valid
even when the K\"ahler metric of $t_3$  is given by a generic
function of $t_2$.
 As for the power $p$ of $t_1$ in the
K\"ahler metric of $t_3$, one can consider two possibilities: \bea
p=3/2 \quad {\rm or} \quad 1.\eea If the local cutoff scale of $T_3$
is given by the string scale, $p=3/2$ would be the correct choice
\cite{Conlon-Pedro}. On the other hand, if the K\"ahler metric of
$T_3$ behaves like those of $\Phi_i$, one would have $p=1$ for which
the local cutoff scale corresponds to the winding mode scale
$M_{*}\sim t_1^{1/4}M_{\rm string}$. Note that $M_{GS}$ takes a
different value depending upon the value of $p$: \bea M_{GS}\,\sim\,
\delta_{GS}M_{\rm string} \,\,\, (p=3/2) \quad \mbox{or}\quad
\delta_{GS}M_{*}\,\,\, (p=1).\eea

 Yukawa couplings of the
 canonically normalized matter fields localized on the MSSM 4-cycle should not have any power-law dependence on the bulk CY volume, and thus
 no power-law dependence on $t_1$.
Under this requirement,
the matter K\"ahler metric is given by
  \bea
 Z_i=\frac{1}{t_1}{\cal
 Y}_i(\hat{t}_2,\hat{t}_3),\eea
where \bea
 \hat{t}_2=t_2-\beta_2\ln t_1, \quad \hat{t}_3=t_3-\beta_3\ln t_1
-\delta_{GS}V_A \eea
 for the moduli mixing parameters $\beta_{2,3}$ which generically
can differ from
 $\alpha_{2,3}$ that describe the moduli mixing in $K_0$.
We can always  choose the normalization convention of $\Phi_i$
 to make ${\cal Y}_i$ have a vacuum
 value of ${\cal O}(1)$, and also take the normalization convention of $T_1$ and $T_3$
for which \bea \label{parameter1}\xi_{\alpha^\prime}={\cal O}(1),
\quad k_{a,A}={\cal O}(1).\eea Then the large volume solution
(\ref{lvs-sol}) (or (\ref{sol-uplifting}) in the presence of
uplifting potential) and the instanton action (\ref{d3action})
indicate that the vacuum value of $t_2$ is of ${\cal O}(1)$ and the
parameter $a={\cal O}(\ln (M_{\rm Planck}/m_{3/2}))$ in our
convention.
 Since the $U(1)_A$-variation of
the gauge kinetic function should be canceled by the loops of
$U(1)_A$-charged fermions, under the assumption that $U(1)_A$
charges of chiral matter fields are generically of ${\cal O}(1)$, we
find $\delta_{GS}={\cal O}(1/8\pi^2)$.  The parameters describing
the loop-induced moduli redefinition, i.e. $\alpha_{2,3}$ and
$\beta_{2,3}$, are expected to have a similar size of ${\cal
O}(1/8\pi^2)$. Summarizing the size of model parameters in our
convention, we have \bea \label{parameter2} \delta_{GS} &\sim&
\alpha_{2,3} \,\sim\, \beta_{2,3} \,=\,{\cal
O}\left(\frac{1}{8\pi^2}\right),\nonumber \\
a &=&{\cal O}\left(\ln\left({M_{\rm
Planck}}/{m_{3/2}}\right)\right)\,=\,{\cal O}(8\pi^2).\eea


 With the K\"ahler potential given in (\ref{model}), the $U(1)_A$
gauge boson mass and the $D$-term are given by \bea \label{u1mass}
M_A^2 &=&
2g_A^2\Big(M_{GS}^2 + M_{PQ}^2\Big), \nonumber \\
D_A &=&
\xi_{FI} +\tilde{M}_{PQ}^2,\eea where \bea \label{u1mass1}
{M^2_{GS}} &=&
\left(\frac{\delta_{GS}}{2}\right)^2\frac{1}{t_1^{p}}, \qquad
{\xi_{FI}}\,=\,
\frac{\delta_{GS}}{2} \frac{\tilde t_3}{t_1^p}, \nonumber
\\
M_{PQ}^2&=&\sum_i
\left(q_i^2Z_i-q_i\delta_{GS}\partial_{t_3}Z_i+\left(\frac{\delta_{GS}}{2}\right)^2\partial^2_{t_3}Z_i\right)
\left\langle \Phi_i^*e^{2q_iV_A}\Phi_i\right\rangle, \nonumber \\
\tilde{M}_{PQ}^2&=&- \sum_i \left(
q_iZ_i-\frac{\delta_{GS}}{2}\partial_{t_3}Z_i\right)\left\langle\Phi_i^*e^{2q_iV_A}\Phi_i\right\rangle.
\eea Here we  use units with $M_{\rm Planck}=1$. The vacuum
values of the gauge-invariant combinations
$\Phi_i^*e^{2q_iV_A}\Phi_i$ will be determined later by the combined
effects of supersymmetry breaking and the $F$-term scalar potential.

Since there is no $D3$ instanton effect of the form $e^{-bT_3}$
\cite{Blumenhagen:2007sm},  the K\"ahler potential and the
superpotential are invariant under the axionic shift of $T_3$: \bea
U(1)_{T_3}:\quad T_3 \rightarrow T_3+\mbox{imaginary constant},\eea
which is explicitly broken by the variation of the holomorphic gauge
kinetic functions. Due to the anomalous $U(1)_A$ gauge symmetry,
this axionic shift symmetry is equivalent  to  the global
$U(1)_{PQ}$ symmetry under which \bea \label{pq} U(1)_{PQ}: \quad
\Phi \rightarrow e^{iq_i \alpha} \Phi_i,\eea where $q_i$ is the
$U(1)_A$ charge of $\Phi_i$, and $\alpha$ is a real constant. The
Green-Schwarz anomaly cancelation \cite{Green-Schwarz} for $U(1)_A$
requires \bea \label{gsanomaly}\frac{1}{2\pi^2} \sum_i q_i {\rm
Tr}(T_a^2(\Phi_i))=k_a \delta_{GS},\quad \frac{1}{2\pi^2}\sum_i
q_i^3 =k_A \delta_{GS},\eea where $k_a$ and $k_A$ are the
coefficients of $T_3$ in the gauge kinetic functions $f_a$ and $f_A$
in (\ref{model}). This means that $U(1)_{PQ}$ is an anomalous global
symmetry which can solve the strong CP problem by the axion
mechanism \cite{pq,axion1,axion2}. Obviously $M_{PQ}\sim
\tilde{M}_{PQ}$
corresponds to the scale where $U(1)_{PQ}$ is spontaneously broken,
and thus to the axion decay constant which is constrained to be
above $10^9$ GeV by astrophysical considerations \cite{axion3}: \bea
M_{PQ}\geq 10^9\,\, {\rm GeV}.\eea

If $t_3$ is stabilized by the $D$-term potential, the  K\"ahler
modulus superfield  $T_3$ becomes a part of the massive $U(1)_A$
vector superfield. As a result, its equation of motion  is encoded
in that of $V_A$, which takes the form \bea \label{supereq}
\frac{\partial K}{\partial V_A}={\cal O}({\cal D}^2\bar{\cal D}^2
V_A),\eea where ${\cal D}^2={\cal D}^\alpha {\cal D}_\alpha$ for the
superspace covariant derivative ${\cal D}_\alpha$. Here the
right-hand-side of the above equation of motion comes from the
variation of the gauge kinetic term in the $N=1$ superspace.
As long as $m_{3/2}\ll M_{GS}$,
this part can be safely ignored in the discussion of moduli
stabilization and SUSY breakdown.
For the K\"ahler potential (\ref{model}), the superfield equation
(\ref{supereq}) is given by \bea \label{supereq1}
4M_{GS}^2\left(V_A-\frac{t_3}{\delta_{GS}}+\frac{\alpha_3}{\delta_{GS}}\ln
t_1\right)+\Phi_i^*e^{2q_iV_A}\Phi_i\left(2q_iZ_i-\delta_{GS}\frac{\partial
Z_i}{\partial \hat{t}_3}\right) = {\cal O}({\cal D}^2\bar{\cal D}^2
V_A).\eea

The K\"ahler potential  in (\ref{model-kahler}) assumes \bea
\tilde{t}_3=t_3-\alpha_3\ln t_1-\delta_{GS}V_A\ll 1.\eea The
superfield equation (\ref{supereq1}) suggests that  this condition
is fulfilled if the matter fields are stabilized as
\bea  \langle Z_i \Phi_i^*e^{2q_i V_A}\Phi_i\rangle \,\sim\,
M_{PQ}^2  \,\ll\, \frac{M_{GS}^2}{\delta_{GS}}.\eea For simplicity,
here we take a stronger condition:
 \bea \label{limit1}M_{PQ}^2 \,\ll\,
M_{GS}^2\eea  since the analysis for such case is rather
straightforward. In the appendix, we will show that similar results
are obtained even when  $M_{PQ}^2\gtrsim M_{GS}^2$ as long as
$M^2_{PQ}\ll M_{GS}^2/\delta_{GS}$.

To determine  the VEVs of $U(1)_A$-charged matter fields  in the
limit $M_{PQ}^2\ll M_{GS}^2$, one can integrate out the massive
$U(1)_A$ vector superfield including $T_3$, and examine the
stabilization of $\Phi_i$ based on the resulting low energy
effective theory \cite{arkani,Choi-Jeong}. For this, it is
convenient to make the field redefinition: \bea
\label{gaugetr}V_A\,\rightarrow\,
V_A+\frac{T_3+T_3^*}{\delta_{GS}},\quad \Phi_i \,\rightarrow\,
e^{-2q_iT_3/\delta_{GS}}\Phi_i,\eea under which the holomorphic
gauge kinetic function is transformed as \bea \label{effective-f}
f_a\,=\,\gamma_a+k_aT_3\quad \rightarrow\quad f^{\rm
eff}_a\,=\,\gamma_a+k_aT_3-\frac{1}{4\pi^2}\sum_i {\rm
Tr}(T_a^2(\Phi_i))\frac{2q_iT_3}{\delta_{GS}}\,=\,\gamma_a,\eea
where we have used  the anomaly matching condition (\ref{gsanomaly})
for the last equality. With this field redefinition, $T_3$ is gauged
away in (\ref{supereq1}), and the solution for the $U(1)_A$  vector
superfield is given by
 \bea
\label{solution}V_A=-\frac{\alpha_3}{\delta_{GS}}\ln t_1
-\left(2q_iZ_i^{\rm eff}-\delta_{GS}\frac{\partial Z_i^{\rm
eff}}{\partial \hat{t_3}}\right)\frac{\Phi_i^*
\Phi_i}{4M_{GS}^2}+{\cal O}\left(\left(\frac{Z_i^{\rm
eff}\Phi_i^*\Phi_i}{M_{GS}^2}\right)^2\right) \eea
where \bea \label{eff-metric} {Z}^{\rm
eff}_i&=&\frac{Z_i(t_1,\hat{t}_2=t_2-\beta_2\ln t_1,
\hat{t}_3=(\alpha_3-\beta_3)\ln
t_1)}{t_1^{2q_i\alpha_3/\delta_{GS}}}\nonumber
\\
&=&
 \frac{{\cal
Y}_i(\hat{t}_2=t_2-\beta_2\ln t_1, \hat{t}_3=(\alpha_3-\beta_3)\ln
t_1)}{t_1^{1+2q_i\alpha_3/\delta_{GS}}}.\eea
The effective K\"ahler potential of light fields can be obtained by
inserting the above solution into the K\"ahler potential
(\ref{model}), which gives  \bea \label{effective-kaehler} K_{\rm
eff}&=&K_0^{\rm eff}(t_1,t_2)+Z_i^{\rm eff}\Phi_i^*\Phi_i
\left(1+{\cal O}\left(\frac{Z_i^{\rm
eff}\Phi_i^*\Phi_i}{M_{GS}^2}\right)\right)\nonumber \\
&=&-3\ln
t_1+\frac{2(\tilde{t}_2^{3/2}-\xi_{\alpha^\prime})}{t_1^{3/2}}+Z_i^{\rm
eff}\Phi_i^*\Phi_i \left(1+{\cal O}\left(\frac{Z_i^{\rm
eff}\Phi_i^*\Phi_i}{M_{GS}^2}\right)\right) .\eea Note that \bea
M_{PQ}^2\,\sim\, \langle Z_i \Phi_i^*e^{2q_i V_A}\Phi_i\rangle
\,\sim \, \langle Z_i^{\rm eff}\Phi^*_i\Phi_i \rangle,\eea and
therefore the above effective K\"ahler potential provides a
controllable description of low energy physics in the limit
$M_{PQ}^2 \ll M_{GS}^2$.

Let us now present a simple model which fixes the axion scale
$M_{PQ}$ by the combined effects of SUSY breaking and a
Planck-scale-suppressed term in the superpotential. This type of
model provides a setup giving the axion scale in the
phenomenologically desirable  range $10^9$ GeV $\leq M_{PQ}\leq$
$10^{12}$ GeV even when the string scale is close to the Planck
scale. To have $M_{PQ}\gtrsim 10^9$ GeV, $U(1)_{PQ}$ should be
broken dominantly by $U(1)_A$-charged but MSSM-singlet matter
fields. As a kind of minimal example, we introduce two such matter
fields, $X$ and $Y$, with the following  K\"ahler potential and
superpotential \cite{murayama-yanagida}: \bea
\label{axion-sector}\Delta
K&=&Z_XX^*e^{2q_XV_A}X+Z_YY^*e^{2q_YV_A}Y,\nonumber
\\
\Delta W&=& \kappa \frac{X^{k+2}Y}{M_{\rm Planck}^k},\eea where
$k=-q_Y/q_X-2$ is a positive
 integer, and $\kappa$ can always be chosen to be a real positive constant which is expected to be of order unity since $M_{\rm
 Planck}$ is the natural scale to determine the coefficients of higher dimensional operators in
the 4D supergravity superpotential.
To determine the vacuum values of $X$ and $Y$, one  can integrate
out $V_A$ and $T_3$ to obtain the effective K\"ahler potential
(\ref{effective-kaehler}), and
 define
 the canonically normalized matter fields
\bea \hat{\Phi}_i=\sqrt{Z_i^{\rm eff}}\Phi_i,\nonumber \eea
 while treating $t_1$ and $t_2$ as a
fixed background.
 Then the effective
potential of $\hat\Phi_i=\{\hat X,\hat Y\}$ takes the form \bea
V_{\rm eff}&=& m_X^2|\hat X|^2 +m_Y^2 |\hat Y|^2 + \left(\hat\kappa
A_\kappa \frac{\hat X^{k+2}\hat
Y}{M_{*}^k}+{\rm h.c.}\right)\nonumber \\
&+& \frac{|\hat\kappa|^2}{M_{*}^{2k}}\left(|\hat{X}|^{2k+4}
+(k+2)^2|\hat{X}|^{2k+2}|\hat Y|^2\right) + {\cal
O}\left(\frac{m_{3/2}^2|\hat \Phi|^4}{M_{GS}^2}\right),  \eea where
$m_i$ ($i=X,Y$) and $A_\kappa$ denote the soft SUSY breaking scalar
mass and $A$-parameter, respectively, and $$\hat\kappa\,=\,
\frac{\kappa}{\sqrt{e^{-(k+3)K_0^{\rm eff}/3}(Z_X^{\rm
eff})^{k+2}Z_Y^{\rm eff}}}\,\simeq\,  \frac{\kappa}{\sqrt{{\cal
Y}_X^{k+2}{\cal Y}_Y}}.$$ Note that $\hat\kappa={\cal O}(\kappa)$ as
${\cal Y}_i$ ($i=X,Y$) are defined to be of order unity, and  the
scalar potential of the canonically normalized matter fields induced
by the Planck-scale-suppressed superpotential  is controlled by the
winding scale:\bea M_{*}\,\sim\, \frac{M_{\rm
Planck}}{t_1^{1/2}}\,\sim\, t_1^{1/4}M_{\rm string}.
\eea  As we will see in the next section, in the limit $M_{PQ}^2 \ll
M_{GS}^2$, the dominant contribution to  $m_i$  comes from the
$U(1)_A$ $D$-term of ${\cal O}(m_{3/2}^2)$, while $A_\kappa$
receives only a  suppressed contribution of ${\cal
O}(m_{3/2}/8\pi^2)$ from the $F$-components of moduli superfields:
\bea m_X^2 &\simeq& -q_X g_A^2D_A \,\simeq\,
-\frac{2q_X\alpha_3}{\delta_{GS}}m_{3/2}^2
\,=\, {\cal
O}(m_{3/2}^2),\nonumber \\
m_Y^2&\simeq&  -q_Y g_A^2 D_A
\,\simeq \,
-(k+2)m_X^2,
\nonumber \\
A_\kappa &=&
{\cal O}\left(\frac{m_{3/2}}{8\pi^2}\right),\eea where we have used
$q_Y = -(k+2)q_X$. A nice feature of the above soft masses is that a
tachyonic $m_X^2$ at tree level can be naturally obtained, which
induces a nonzero vacuum value of $X$ at an intermediate
scale. One just needs to assume  $q_X\alpha_3/\delta_{GS}>0$, giving
$m_X^2 <0$.
We then find that $\hat X$ and $\hat Y$ are stabilized at the
following vacuum values: \bea\label{vev_PQ} \langle \hat X\rangle
&=& \left(\frac{m_XM_*^k}{\hat{\kappa}\sqrt{k+2}}\right)^{1/(k+1)}
\left(1+ {\cal O}
 \left(\frac{A_\kappa^2}{m_{3/2}^2}\right)\right),
 \nonumber\\
\langle \hat Y\rangle &=&\frac{A_{\kappa}}{2(k+2)m_Y}\langle\hat X\rangle
\left(1+ {\cal O}\left(\frac{\langle \hat X\hat
X^*\rangle}{M_{GS}^2}\right)\right),  \eea which give
 \bea \label{vev_PQ1}&&\langle \hat X\rangle \,\sim\,  M_{PQ}\,\sim\, \frac{(m_{3/2} M_{\rm
Planck}^k)^{1/(k+1)}}{t_1^{k/2(k+1)}}, \nonumber \\
&& \langle \hat Y\rangle \,\sim\, \left(\frac{A_\kappa}{m_Y} \right)
\langle \hat X\rangle\, \sim\, {\cal O}\left( \frac{\langle \hat
X\rangle}{8\pi^2}\right).\eea
With this result, one can choose appropriate values of model
parameters to get an intermediate axion scale in the range $10^9$
GeV $\leq M_{PQ}\leq$ $10^{12}$ GeV.

In the above, we have presented a simple model in which  the axion
scale $M_{PQ}$ is determined  by the combined effect of
supersymmetry breaking and a Planck-scale suppressed term in the
superpotential.
For simplicity, we assumed that the axion scale is lower than the
$U(1)_A$ vector boson mass as  \bea M_{PQ} \ll M_{GS} \sim
\delta_{GS} t_1^{\frac{3-2p}{4}} M_{\rm string},\eea where $p$
($=3/2$ or 1) denotes the power of $t_1$ in the K\"ahler potential
of the MSSM cycle modulus $T_3$.
 On the other hand, in LVS
 we generically have $M_{\rm string}/M_{\rm Planck}
 \sim 1/t_1^{3/4}$ and $m_{3/2}/M_{\rm string}\sim W_0/t_1^{3/4}$,
 yielding \bea
M_{GS}\,\sim\, \delta_{GS}M_{\rm
Planck}\left(\frac{1}{W_0}\frac{m_{3/2}}{M_{\rm
Planck}}\right)^{p/3}\sim\, \Big(10^{16}-10^{17}\Big)\times
\left(\frac{1}{W_0}\frac{m_{3/2}}{M_{\rm Planck}}\right)^{p/3}
\,\,{\rm GeV}, \eea
 where $W_0$ is the
 flux-induced
constant in the effective superpotential.
As was noticed in \cite{Conlon-Pedro}, in the presence of
loop-induced moduli-mixing, the MSSM gauginos get a mass of ${\cal
O}(m_{3/2}/8\pi^2)$, and therefore $m_{3/2}$ is required to be in
multi-TeV range, e.g. $m_{3/2}={\cal O}(10)$ TeV,  in order to
realize the weak scale SUSY scenario. If we further assume
$W_0\sim{\cal O}(1)$ and $p=3/2$,
the resulting $M_{GS}$ is in the range of $10^9-10^{10}$ GeV, and
therefore  might not be high enough to assure the condition
$M_{GS}\gg M_{PQ}> 10^9$ GeV which has been used throughout our
analysis.  In such a case, one might need to stabilize the matter
fields at a point giving $M_{PQ}$ comparable to or even higher than
$M_{GS}$, which would require a separate analysis. In the appendix,
we show that it is also possible to stabilize $T_3, X$ and $Y$
within the model of (\ref{model}) and (\ref{axion-sector}) at a
point giving $\langle \hat{X}\rangle \sim M_{PQ} \gtrsim M_{GS}$, as
long as $\langle \hat X^* \hat X\rangle \ll M_{GS}^2/\delta_{GS}$,
and the resulting SUSY breaking patterns are similar to those in the
case of $\langle \hat{X}^*\hat X\rangle \ll M_{GS}^2$. In such
situation, $M_{GS}$ corresponds to the QCD axion scale, while
$M_{PQ}$ determines the $U(1)_A$ gauge boson mass.

\section{SUSY breakdown and soft terms}
\label{moduli mediation}

We are now ready to  discuss SUSY breakdown and the soft terms in
the MSSM sector. In the LVS models, the $F$-components of the
K\"ahler moduli $T_i$ ($i=1,2,3$) and the $D$-component of the
$U(1)_A$ vector superfield $V_A$ are the prime candidates for the
origin of soft terms
\cite{moduli-mediation,d-term-mediation1,d-term-mediation2}. If
there exist gauge-charged messengers which have a Yukawa coupling to
$X$ and/or $Y$, there can be gauge-mediated soft terms arising from
the $F$-components of $X$ and/or $Y$ \cite{gauge-mediation}. We
first evaluate the vacuum values of these SUSY breaking auxiliary
components using the results of the previous two sections.

From the moduli vacuum values of $T_{1}$ and $T_2$ in
(\ref{lvs-sol}), it is straightforward to find
 \bea\label{Ft1t2} &&\frac{F^{T_1}}{t_1}= m_{3/2}\left[1+ {\cal O}
\left(\frac{\xi_{\alpha'}}{a\tilde t_2
t_1^{3/2}}\right)\right], \nonumber\\
&&\frac{F^{T_2}}{\tilde{t}_2}=  m_{3/2}\left[\frac{3 }{2a\tilde t_2}
+ {\cal O}\left(\frac{1}{(a\tilde t_2)^{2}}\right)\right], \eea
 where the $F$-component of a generic chiral superfield $\Phi^I$ is defined as
$F^I =-e^{K/2}K^{I\bar{J}}(D_JW)^*$, and this expression of
$F^{T_{i}}$ ($i=1,2$) is not affected by the uplifting potential.
Combining the vacuum values   of the PQ sector fields in
(\ref{vev_PQ}) with the superfield equation (\ref{supereq1}),  we
also find the following vacuum configuration of the  $U(1)_A$ sector
fields: \bea \label{vev_U1A} && \hskip -1.5cm \langle \hat X^*\hat
X\rangle\,\equiv\, \langle Z_X X^* e^{2q_XV_A} X\rangle  \,\sim\,
M_{PQ}^2,\nonumber \\
&& \hskip -1.5cm\langle \hat Y^* \hat Y\rangle   \,\equiv \, \langle
Z_Y Y^* e^{2q_YV_A} Y\rangle
\,\sim\, \frac{A_\kappa^2 M_{PQ}^2}{m_Y^2}\,\sim\, \frac{M_{PQ}^2}{(8\pi^2)^2}, \nonumber \\
 \tilde{t}_3 &=& t_3-\alpha_3\ln t_1-\delta_{GS}V_A\,=\, {\cal
O}\left(\frac{\delta_{GS}M_{PQ}^2}{M_{GS}^2}\right),\nonumber
\\
\frac{F^X}{X}&\sim &
A_\kappa\,\sim\,\frac{m_{3/2}}{8\pi^2}\left[1+{\cal
O}\left(\frac{M_{PQ}^2}{M_{GS}^2}\right)\right]\nonumber \\
 \frac{F^Y}{Y}&\sim& \frac{m_Y^2}{A_\kappa}\,\sim\, \frac{D_A}{A_\kappa}\,\sim\, 8\pi^2
m_{3/2},
\nonumber \\
F^{T_3}&=&\alpha_3 \frac{F^{T_1}}{t_1} +{\cal
O}\left(\frac{m_{3/2}\delta_{GS}M_{PQ}^2}{M_{GS}^2}\right) =
\alpha_3 m_{3/2} + {\cal
O}\left(\frac{m_{3/2}\delta_{GS}M_{PQ}^2}{M_{GS}^2}\right),\nonumber \\
g_A^2D_A&=&
\frac{2\alpha_3}{\delta_{GS}}\left|\frac{F^{T_1}}{t_1}\right|^2
+{\cal O}\left(\frac{m_{3/2}^2M_{PQ}^2}{M_{GS}^2}\right)
=\frac{2\alpha_3m_{3/2}^2}{\delta_{GS}}+{\cal
O}\left(\frac{m_{3/2}^2M_{PQ}^2}{M_{GS}^2}\right),\eea where $m^2_Y$
denotes the soft scalar mass of $Y$, which is of the order of $D_A$.
Note that some vacuum values, for instance those of
$F^{\Phi_i}/\Phi_i$ ($\Phi_i=X,Y$) and the scalar and $F$ components
of $V_A$,  are not invariant under the $T_3$-dependent field
redefinition (\ref{gaugetr}). More specifically, the original $V_A$
in (\ref{supereq1}) is defined in the Wess-Zumino gauge, and
therefore has vanishing scalar and $F$ components. On the other
hand,  $V_A$ in (\ref{solution}) after the field redefinition
(\ref{gaugetr}) is defined  in the unitary gauge containing the
Goldstone superfield $\propto T_3+T_3^*$, and therefore has nonzero
scalar and $F$ components coming from $T_3+T^*_3$. As for
$F^{\Phi_i}/\Phi_i$ ($\Phi_i=X,Y$), the above results denote the
vacuum values before the redefinition (\ref{gaugetr}).

Although the above vacuum values of  the $U(1)_A$ sector fields have
been derived in the limit $M_{PQ}^2 \ll M_{GS}^2$, we find (as
explained in the appendix) that they remain to be valid even for
$M_{PQ}^2\gtrsim M_{GS}^2$, at least qualitatively, as along as
$\tilde t_3 \sim \delta_{GS}M_{PQ}^2/ M_{GS}^2\ll 1$, which would be
required for the K\"ahler potential to be expanded in powers of
$\tilde t_3$ as in (\ref{model-kahler}).

 A notable feature of LVS is the no-scale
structure of the large volume modulus $t_1$, which leads to a strong
suppression of the anomaly mediated contributions.
Anomaly mediation is described
most conveniently by the super-Weyl-invariant compensator
formulation of 4D SUGRA, involving a chiral compensator superfield
$C$ \cite{anomaly-mediation}. One can then choose a super-Weyl gauge
in which the SUSY breaking (but Poincare-invariant) component of
ordinary SUGRA multiplet is vanishing, e.g. \bea \left.{\cal
R}\right|_{\theta=\bar\theta=0}=0,\eea where ${\cal R}$ is the
chiral curvature superfield in $N=1$ superspace.  One could also
choose the Einstein frame gauge for which
$C_0\equiv\left.C\right|_{\theta=\bar\theta=0}=e^{K/6}$. Thus
the $F$-component of the compensator superfield is given by \bea
\frac{F^C}{C_0}= m_{3/2}+\frac{1}{3}K_IF^I.\eea
Using the results on the  moduli and matter $F$-components, we find
\bea \frac{F^C}{C_0}={\cal O}\left(\frac{m_{3/2}M_{\rm
string}^2}{M_{\rm Planck}^2},\frac{m_{3/2}M_{PQ}^2}{M_{\rm
Planck}^2}\right),\eea where
the piece  of  ${\cal O}(m_{3/2}M_{PQ}^2/M_{\rm Planck}^2)$
originates from the $K_{T_3}F^{T_3}$, while the other piece of
${\cal O}(m_{3/2}M_{\rm string}^2/M_{\rm Planck}^2)$ can arise from
the mixing between $T_1$ and the string dilaton that appears in the
$\alpha^\prime$ correction to the K\"ahler potential.  This assures
that anomaly mediation in LVS can be safely ignored.


With the SUSY breaking auxiliary components given above, we can
compute the soft terms of the MSSM gauge and matter multiplets as
well as those of the PQ sector matter multiplets. Here we will focus
on the soft terms induced  by the moduli $F$-components
\cite{moduli-mediation} and the $U(1)_A$ $D$-component
\cite{d-term-mediation1,d-term-mediation2}, although there can be
gauge-mediated soft terms as well. For instance, if there exist
gauge-charged messenger fields $\Phi+\Phi^c$ with a Yukawa coupling
$\propto Y\Phi\Phi^c$, gaugino and scalar masses of ${\cal
O}(m_{3/2})$ can be induced by $F^{Y}/Y\sim 8\pi^2 m_{3/2}$ through
the conventional gauge mediation mechanism \cite{gauge-mediation},
and these gauge-mediated gaugino masses will dominate over the
moduli-mediated gaugino masses of ${\cal O}(m_{3/2}/8\pi^2)$.
However the presence of such gauge mediation is strongly model
dependent. Particularly it depends on whether the model contains
exotic matter fields with the right quantum numbers and right
couplings. In any case, it is straightforward to incorporate the
gauge-mediated soft terms, if there exist any, with the more generic
moduli-mediated or $D$-term induced soft terms on which we will
concentrate in the following.

To evaluate the moduli-mediated  and $D$-term induced soft
parameters, let
 $T_I$ denote the SUSY breaking moduli
superfields (not including the compensator $C$), and $\Phi_i$ denote
the visible sector chiral superfields with canonically normalized
scalar components $\hat\phi_i$.
 For a generic 4D SUGRA
model described by \bea K&=& K_0(T_I,
T_{I}^*)+Z_i(T_I,T_I^{*})\Phi_{i}^*e^{2q_i V_A}\Phi_i, \nonumber
\\
W&=&
\tilde W_0(T_I)+\frac{1}{3!}\lambda_{ijk}(T_I)\Phi_i\Phi_j\Phi_k+\frac{1}{n!}\kappa_{i_1i_2..i_n}(T_I)
\Phi_{i_1}\Phi_{i_2}..\Phi_{i_n},
\nonumber \\
f_a&=&f_a(T_I),\eea soft  SUSY breaking terms of canonically
normalized components  fields take the form \bea {\cal L}_{\rm
soft}&=&-\frac{1}{2}M_a\lambda^a\lambda^a-\frac{1}{2}m_i^2|\hat\phi_i|^2-\frac{1}{3!}A_{ijk}\hat\lambda_{ijk}\hat\phi_i\hat\phi_j\hat\phi_k
\nonumber \\
&&- \frac{1}{n!}A_\kappa^{i_1i_2..i_n}
\hat{\kappa}_{i_1i_2..i_n}\hat\phi_{i_1}\hat\phi_{i_2}..\hat\phi_{i_n}
+{\rm h.c.}, \eea where  $\hat{\lambda}_{ijk}$ and
$\hat{\kappa}_{i_1i_2..i_n}$ denote  the canonically normalized
Yukawa couplings,
 \bea
\hat{\lambda}_{ijk}&=&\frac{\lambda_{ijk}}{\sqrt{e^{-K_0}Z_iZ_jZ_k}},\nonumber
\\
\hat{\kappa}_{i_1i_2..i_n}&=&
\frac{\kappa_{i_1i_2..i_n}}{\sqrt{e^{-nK_0/3}Z_{i_1}Z_{i_2}..Z_{i_n}}},\nonumber\eea
and the soft SUSY breaking parameters (at scales around the cutoff
scale) are then given by \bea \label{soft-masses}A_{ijk}&=&-F^I\partial_I
\ln \left(\frac{\lambda_{ijk}}{e^{-K_0}Z_iZ_jZ_k}\right),\nonumber
\\
A_{\kappa}^{i_1i_2..i_n}&=& (n-3)\frac{F^C}{C_0}-F^I\partial_I \ln
\left(\frac{\kappa_{i_1i_2..i_n}}{e^{-nK_0/3}Z_{i_1}Z_{i_2}..Z_{i_n}}\right),\nonumber
\\
m_i^2 &=&
\frac{2}{3}V_F-F^IF^{\bar{J}}\partial_I\partial_{\bar{J}}\ln
\left(e^{-K_0/3} Z_i\right)-\left(q_i+\eta^I\partial_I\ln
Z_i\right)g_A^2 D_A\nonumber \\
\frac{M_a}{g_a^2}&=&\frac{1}{2}F^I\partial_I f_a
-\frac{1}{8\pi^2}\sum_i {\rm Tr}(T^2_a(\Phi_i))F^I\partial_I\ln
(e^{-K_0/3}Z_i)+\frac{b_a}{16\pi^2}\frac{F^C}{C_0}, \eea where
$V_F=K_{I\bar{J}}F^IF^{\bar{J}}-3m_{3/2}^2$ is the $F$-term scalar
potential. Here we consider the tree level contributions (in the
sense of 4D effective SUGRA) to $m_i$ and $A$-parameters as they
provide the dominant part at the cutoff scale.   On the other hand,
we included the full 1-loop contributions to gaugino masses
\cite{gaugino-loop}, which can be relevant for the gaugino masses
derived from  the effective theory after the massive $U(1)_A$ vector
multiplet (including $T_3$) is integrated out.

 The above
expression of soft masses can be applied to the LVS model
(\ref{model}) with the PQ sector given  by (\ref{axion-sector}).
We find
 \bea
 m_i^2 &\simeq & -q_i g_A^2 D_A \,=\,\left[-\frac{2q_i\alpha_3}{\delta_{GS}}
 +{\cal O}\left(\frac{M_{PQ}^2}{M_{GS}^2}\right)\right]m_{3/2}^2
 \,=\,{\cal O}(m_{3/2}^2),
 \nonumber \\
 A_{ijk}
 &\simeq &\left[\left(\frac{3}{2a}-\beta_2\right)\partial_{\hat t_2}
 \ln{\cal Y}_i{\cal Y}_j{\cal Y}_k + (\alpha_3-\beta_3)\partial_{\hat t_3}
 \ln{\cal Y}_i{\cal Y}_j{\cal Y}_k\right]m_{3/2} ={\cal O}\left(\frac{m_{3/2}}{8\pi^2}\right)
 \nonumber\\
A_\kappa &\simeq
&\left[\left(\frac{3}{2a}-\beta_2\right)\partial_{\hat t_2}
 \ln({\cal Y}_X^{k+2}{\cal Y}_Y)+ (\alpha_3-\beta_3)\partial_{\hat t_3}
 \ln({\cal Y}_X^{k+2}{\cal Y}_Y)\right]m_{3/2}
={\cal O}\left(\frac{m_{3/2}}{8\pi^2}\right) \nonumber\\
\frac{M_a}{g_a^2} &\simeq & \frac{k_a}{2}F^{T_3} \,=\,
\left[\frac{\alpha_3 k_a}{2}+{\cal
O}\left(\frac{\delta_{GS}M_{PQ}^2}{M_{GS}^2}\right)\right]m_{3/2}
\,=\,{\cal O}\left(\frac{m_{3/2}}{8\pi^2}\right), \eea where
 we have kept only
the dominant part in the limit $M_{PQ}^2 \ll M_{GS}^2$. Here
 the matter K\"ahler metric is given by  $Z_i={\cal Y}_i(\hat t_2, \hat t_3)/t_1$, and we assumed
 that ${\cal Y}_i$ is a
generic function of  $\hat t_2 =t_2-\beta_2\ln t_1$ and $\hat t_3
=t_3-\beta_3\ln t_1-\delta_{GS}V_A$, with $\partial_a \ln {\cal
Y}_i={\cal O}(1)$ ($a=\hat t_2, \hat t_3$).

The most notable feature of the above soft masses is the relative
enhancement of the scalar masses compared to the gaugino masses.
Gaugino masses and $A$-parameters induced by the moduli
$F$-components are generically of ${\cal O}(m_{3/2}/8\pi^2)$, while
sfermion masses induced by the $U(1)_A$ $D$-term
 are of ${\cal O}(m_{3/2})$.
Note that loop-induced moduli mixing, particularly the one
described by $\alpha_3$, is crucial for the soft masses comparable
to $m_{3/2}$ or $m_{3/2}/8\pi^2$.
An intriguing feature of our results is that the $D$-term induced
sfermion masses  are ${\cal O}(m_{3/2})$, although they result from
a loop-induced moduli mixing. This is due to the suppression of the
$U(1)_A$ gauge boson mass-square by $\delta_{GS}^2$. Since the
$D$-term contribution to sfermion masses arises from the exchange of
the $U(1)_A$ gauge boson, the loop-suppression factor $\alpha_3$ is
compensated by the enhancement factor $1/\delta_{GS}$ in $1/M_A^2$,
and this makes the $D$-term contribution to be of ${\cal
O}(m_{3/2}^2)$.

In the case $M_{PQ}^2\ll M_{GS}^2$, the above soft parameters  can
be obtained also from the effective theory constructed  by
integrating out the massive $V_A$ and $T_3$. Indeed, by applying
(\ref{soft-masses}) to the effective theory described by the
effective K\"ahler potential (\ref{effective-kaehler}) and the
effective gauge kinetic function (\ref{effective-f}), we find the
same result up to small corrections suppressed by
$M_{PQ}^2/M_{GS}^2$:
 \bea
 m_i^2 &=&
 -|F^{T_1}|^2\partial_{T_1}\partial_{T_1^*}\ln (e^{-K_0^{\rm
 eff}/3}Z_i^{\rm eff}) \,=\,-\frac{2q_i\alpha_3}{\delta_{GS}} m_{3/2}^2,
 \nonumber \\
 A_{ijk}&=&
 \left[\frac{3}{2a}\partial_{t_2}
 \ln{\cal Y}^{\rm eff}_i{\cal Y}^{\rm eff}_j{\cal Y}^{\rm eff}_k + \partial_{t_1}
 \ln{\cal Y}^{\rm eff}_i{\cal Y}^{\rm eff}_j{\cal Y}^{\rm
 eff}_k\right]m_{3/2},
 \nonumber\\
A_\kappa  &=&\left[\frac{3}{2a}\partial_{t_2}
 \ln\Big(({\cal Y}^{\rm eff}_X)^{k+2}{\cal Y}^{\rm eff}_Y\Big)+t_1 \partial_{t_1}
 \ln\Big(({\cal Y}^{\rm eff}_X)^{k+2}{\cal Y}^{\rm
 eff}_Y\Big)\right]m_{3/2},
\nonumber\\
\frac{M_a}{g_a^2} &=&  -\frac{1}{8\pi^2}\sum_i {\rm
Tr}(T^2_a(\Phi_i))F^{T_1}\partial_{t_1}\ln (e^{-K^{\rm
eff}_0/3}Z^{\rm eff}_i)  \,=\,\frac{\alpha_3 k_a}{2} m_{3/2},  \eea
where \bea {Z}^{\rm
eff}_i(t_1,t_2)&=&\frac{Z_i(t_1,\hat{t}_2=t_2-\beta_2\ln t_1,
\hat{t}_3=(\alpha_3-\beta_3)\ln
t_1)}{t_1^{2q_i\alpha_3/\delta_{GS}}}\nonumber
\\
&=&
 \frac{{\cal
Y}_i(\hat{t}_2=t_2-\beta_2\ln t_1, \hat{t}_3=(\alpha_3-\beta_3)\ln
t_1)}{t_1^{1+2q_i\alpha_3/\delta_{GS}}}\nonumber \\
&\equiv& \frac{{\cal Y}^{\rm
eff}_i(t_1,t_2)}{t_1^{1+2q_i\alpha_3/\delta_{GS}}},\nonumber \eea
and we have used the anomaly matching condition (\ref{gsanomaly})
for the gaugino masses.



\section{conclusions}

As we have seen, the LVS scenario \cite{LVS} leads to a somewhat
different pattern of soft terms than the conventional KKLT scenario
\cite{KKLT}. While in the LVS case we find soft scalar masses of the
order of the gravitino mass (due to the $D$-term contribution), the
KKLT scenario leads to suppressed scalar masses of order
$m_{3/2}/\ln(M_{\rm Planck}/m_{3/2})$ as expected in mirage
mediation \cite{mirage1a,mirage2}. The moduli-mediated gaugino
masses and $A$-terms appear to be of order $m_{3/2}/\ln(M_{\rm
Planck}/m_{3/2})$ in both cases. Therefore, unless there exist more
model dependent gauge-mediated contributions dominating over the
moduli mediation, soft masses in LVS with moduli-mixing shows a
loop-hierarchy pattern: $m_i\sim 8\pi^2 M_a$. Another key difference
between LVS and KKLT is the relative importance of anomaly
mediation. In the LVS, due to the no-scale structure, anomaly
mediation is negligible, while in the KKLT scenario it becomes
comparable to the moduli-mediation, leading to the mirage
unification of soft masses at a scale different from $M_{GUT}\sim
2\times 10^{16}$ GeV \cite{mirage2}.

The two set-ups differ also in the way to obtain a small value of
$m_{3/2}$. In the KKLT scenario this comes from a small value of the
superpotential $W$, while in the LVS models one assumes $W\sim 1$
and a small $m_{3/2}$ is the result of the large volume suppression.
Large volume, of course, implies a large hierarchy between the
Planck scale and the string scale ${M^2_{\rm Planck}}/{M^2_{\rm
string}}\,\sim\, {\cal V}_{CY}$. The natural setting for LVS would
be a string scale at an intermediate value of approximately
$10^{11}$ GeV and a gravitino mass as well as the soft mass terms at
the (multi) TeV-scale. In the KKLT set-up the natural value of
$M_{\rm string}$ would be rather large, somewhere between the Planck
scale and a possible GUT scale at $10^{16}$ GeV. The two scenarii
can be connected in principle by changing the value of the
superpotential. With the results for the soft terms in the presence
of loop-induced moduli mixing, assuming soft masses to be in the TeV
range, it is likely that a gravitino mass much heavier than ${\cal
O}(8\pi^2)$ TeV requires a severe fine tuning of K\"ahler potential.
For the LVS case (assuming $W\sim 1$) this would imply the string
scale is around $10^{11}$ GeV, which is quite small compared to the
GUT scale of $10^{16}$ GeV. In that sense, the pure LVS-scenario
might be difficult to be compatible with gauge coupling unification
at $10^{16}$ GeV. Such a GUT-picture would require a value of the
superpotential that is small compared to one.

Our considerations are valid for the simple set-up described and we
have focused on the generic contributions to soft terms from the
moduli $F$-components and the $U(1)_A$ $D$-term. Other more model
dependent contributions could be present as well. For example, the
fields $X$ and $Y$ introduced in section 3 to break the global
Peccei-Quinn symmetry at an intermediate scale could act as
the origin of additional ``gauge mediated'' contributions. It is
straightforward to incorporate those gauge mediated contributions
with the soft terms discussed in this paper. This, in connection
with the incorporation of a QCD axion, will be discussed in detail
in a future publication \cite{future}.

\section*{Acknowledgments}

KC and CSS are supported by the KRF Grants funded by the Korean
Government (KRF-2008-314-C00064 and KRF-2007-341-C00010) and the
KOSEF Grant funded by the Korean Government (No. 2009-0080844). HPN
and MT are supported by the SFB-Tansregio TR33 "The Dark Universe"
(Deutsche Forschungsgemeinschaft) and the European Union 7th network
program "Unification in the LHC era" (PITN-GA-2009-237920).

\appendix

\section{}

In section 3, we have discussed the stabilization of the $D$-flat
direction based on the effective theory constructed in the limit
$M_{PQ}^2\ll M_{GS}^2$. Here we provide an analysis of the
stabilization of the $U(1)_A$ sector in more general situation
including the case $M_{PQ}\sim M_{GS}$.

Under the assumption that $\tilde{t}_3=t_3-\alpha_3\ln
t_1-\delta_{GS}V_A$ has a small vacuum value, which would be
fulfilled if $M_{PQ}^2 \ll M_{GS}^2/\delta_{GS}$, the K\"{a}her
potential and superpotential of the model are given by
\bea\label{model-appen} K&=&-3\ln t_1
+\frac{2(\tilde{t}_2^{3/2}-\xi_{\alpha^\prime})}{t_1^{3/2}}+
\frac{1}{2t_1^p}\Big(\tilde t_3^2 + {\cal O}(\tilde t_3^3)\Big)
\nonumber \\
&&+\,
Z_XX^* e^{2q_X V_A}X+Z_YY^*e^{2q_YV_A}Y,   \nonumber\\
W&=&W_0 + A e^{-aT_2} + \kappa\frac{X^{k+2}Y}{M_{\rm
Planck}^k},\quad \eea where $Z_i ={\cal Y}_i(\hat t_2,\hat
t_3)/{t_1}$ ($i=X,Y$). Inserting the values of $t_{1,2}$ and
$F^{T_{1,2}}$ obtained in (\ref{lvs-sol}) and (\ref{Ft1t2}) into the
SUGRA scalar potential:
$$ V_{\rm SUGRA}=
e^K\left\{K^{I\bar{J}}D_IW(D_JW)^*-3|W|^2\right\}+\frac{1}{2}g_A^2D_A^2,$$
 we find
the following scalar potential of $t_3,X$ and $Y$:
\bea
\label{u1a-potential}V(t_3,X,Y)&=&\frac{g_A^2}{2}\left(M_{GS}^2\frac{2\tilde
t_3}{\delta_{GS}}
- q_X\tilde Z_X|X|^2-q_Y\tilde Z_Y|Y|^2\right)^2\nonumber\\
&&-\ \left(\frac{4m_{3/2}^2M_{GS}^2}
{\delta_{GS}^2}\right)\left(\alpha_3\tilde t_3
+\frac{(2-p)(p-1)}{2}\tilde t_3^2\right)\nonumber\\
&&-\ m_{3/2}^2\Big(\beta_a\partial_a\ln{\cal Y}_X + (\gamma_a
\gamma_b\partial_a\partial_b\ln{\cal Y}_X)\Big)
Z_X|X|^2\nonumber\\
&&-\ m_{3/2}^2\Big(\beta_a\partial_a\ln{\cal Y}_Y + (\gamma_a
\gamma_b\partial_a\partial_b\ln{\cal Y}_Y)\Big)
Z_Y|Y|^2\nonumber\\
&&+\ \left(A_{\kappa}\frac{\kappa X^{k+2}Y}{t_1^{3/2}}+h.c.\right) +
\frac{|\kappa|^2}{t_1^3}\left(\frac{|(k+2)X^{k+1}Y|^2}{Z_X}
+\frac{|X^{k+2}|^2}{Z_Y}\right),\eea where \bea\label{parameter}
q_i\tilde{Z}_i&=&\left(q_i-\frac{\delta_{GS}}{2}\partial_{\hat
t_3}\right)Z_i =q_i Z_i -\frac{\delta_{GS}}{2}\partial_{\hat
t_3}Z_i,\nonumber
\\
\beta_a\partial_a &=& \beta_2\partial_{\hat t_2} + \beta_3\partial_{\hat t_3},\nonumber\\
\gamma_a\partial_a&= &
\left(\frac{3}{2a}-\beta_2\right)\partial_{\hat t_2} +
\Big(\alpha_3-\beta_3+(p-1)\tilde t_3\Big)\partial_{\hat t_3},
\nonumber
\\A_{\kappa} &=&\left(\gamma_a\partial_a \ln({\cal Y}_X^{k+2}{\cal Y}_Y)\right) m_{3/2}.
\eea From the SUGRA expression of the $F$-component, i.e.
$F^I=-e^{K/2}K^{I\bar{J}}(D_JW)^*$,  we find also
 \bea F^{T_3} &=&
\Big(\alpha_3 +(p-1)\tilde t_3\Big)m_{3/2}\nonumber\\
\frac{F^{\Phi_i}}{\Phi_i} &=&  - \left(\gamma_a\partial_a \ln {\cal
Y}_i\right)m_{3/2} -\frac{(\partial_{ \Phi_i} W_{\rm
m})^*}{t_1^{3/2} Z_i\Phi_i} \quad (\Phi_i=X,Y),\eea where $W_{\rm
m}=\kappa X^{k+2}Y/M_{\rm Planck}^k$. The stationary point of  the
potential (\ref{u1a-potential}) is given by
 \bea
 \label{u1a-solution}
 \tilde{t}_3 &=& \frac{\delta_{GS} M_{PQ}^2}{M_{GS}^2}
 \left(1+{\cal O}(\delta_{GS})\right),\nonumber \\
 |\hat X|& = &\sqrt{Z_X|X|^2}= \left(\sqrt{\frac{q_X g_A^2 D_A}{k+2}}\,
\frac{M_*^{k}}{|\hat\kappa|}\right)^{1/(k+1)}
\left(1+ {\cal O}\left(\frac{A_\kappa^2}{D_A}\right)\right),\nonumber\\
|\hat Y|&=&\sqrt{Z_Y|Y|^2}=\frac{|A_\kappa \hat X|}{2\sqrt{q_Xg_A^2
D_A(k+2)^3}} \left(1+ {\cal
O}\left(\frac{A_\kappa^2}{D_A}\right)\right),\eea where \bea
M_*&=&\frac{M_{\rm Planck}}{t_1^{1/2}}\,\sim\, t_1^{1/4}M_{\rm string},\nonumber\\
M_{PQ}^2&=& \left(q_X-\frac{\delta_{GS}}{2}\partial_{\hat
t_3}\right)^2Z_X|X|^2
+ \left(q_X-\frac{\delta_{GS}}{2}\partial_{\hat t_3}\right)^2Z_Y|Y|^2,\nonumber\\
g_A^2 D_A &= & g_A^2\left(\frac{2M_{GS}^2 \tilde t_3 }{\delta_{GS}}
-q_i\tilde Z_i |\Phi_i|^2\right)=
\left(\frac{2q_X\alpha_3}{\delta_{GS}}+ (2-p)(p-1)
\frac{M_{PQ}^2}{M_{GS}^2}\right)\frac{m_{3/2}^2}{q_X} . \nonumber
\eea To see that this stationary  point is a stable (local) minimum,
we can compute the mass-square eigenvalues $m_{1,2}^2$ of the
$D$-flat directions. We then find \bea
m_{1}^2 &=& \left(\frac{4(k+1)\left(\frac{2q_X\alpha_3}
{\delta_{GS}}\right) +k\frac{M_{PQ}^2}{M_{GS}^2}}{1 + \frac{M_{PQ}^2}{M_{GS}^2}}\right)m_{3/2}^2\\
m_{2}^2 &=& -2q_Y g_A^2 D_A=
2(k+2)\left(\frac{2q_X\alpha_3}{\delta_{GS}}+ (2-p)(p-1)
\frac{M_{PQ}^2}{M_{GS}^2}\right)m_{3/2}^2, \eea showing that both
eigenvalues are positive, so the solution (\ref{u1a-solution}) is
indeed a stable (local) minimum. We can now make an order of
magnitude estimate for the vacuum configuration of the $U(1)_A$
sector fields: \bea D_A &\sim & \left(\frac{\alpha_3}{\delta_{GS}}+
\frac{(p-1)M_{PQ}^2}{M_{GS}^2}\right)m_{3/2}^2,\nonumber \\
F^{T_3}&\sim& \left(\alpha_3
+\frac{(p-1)\delta_{GS}M_{PQ}^2}{M_{GS}^2}\right)m_{3/2},
\nonumber\\
A_\kappa&\sim& \left(\frac{1}{8\pi^2}+
\frac{(p-1)\delta_{GS}M_{PQ}^2}{M_{GS}^2}\right)m_{3/2} \nonumber \\
|\hat X| & \sim & M_{PQ}\sim \left(\sqrt{D_A}
M_*^k\right)^{1/(k+1)},
\quad
 |\hat Y| \,\sim \,\frac{A_\kappa}{\sqrt{D_A}}
|\hat X|,
\nonumber\\
\frac{F^X}{X} &\sim & (\gamma_a\partial_a\ln{\cal Y}_X)m_{3/2}\sim
A_\kappa,\quad \frac{F^Y}{Y} \,\sim\,
\frac{D_A}{A_\kappa},\eea where $\delta_{GS}\sim \alpha_3\sim
1/8\pi^2$, and we have ignored the coefficients of order unity in
the expression.
 Note that in the limit $M_{PQ}^2 \ll
M_{GS}^2$, the above results reproduce (\ref{vev_PQ}) and
(\ref{vev_U1A}) obtained in section 4 based on the effective theory
constructed by integrating out the massive $U(1)_A$ vector
superfield in the limit $M_{PQ}^2\ll M_{GS}^2$. The above results
show also that (\ref{vev_PQ}) and (\ref{vev_U1A}) are valid even
when $M_{PQ}^2\gtrsim M_{GS}^2$, at least qualitatively, as along as
$\tilde t_3\sim \delta_{GS}M_{PQ}^2/M_{GS}^2\ll 1$, which is
required for the K\"ahler potential to have a meaningful expansion
in powers of $\tilde{t}_3$ as in (\ref{model-appen}).

\end{document}